\documentclass[fleqn,a4paper]{article}
\usepackage{latexsym}
\usepackage{ifthen}
\usepackage[dvips]{graphicx}

\begin{document}

\title{A Solution to the Quantum Measurement Problem}
\author{Andrew Gray \\ London, UK \\ agray592@aol.com}
\date{August 8, 2004}

\maketitle      

\begin{abstract}
  A new formulation of quantum mechanics is developed which does not require the concept of the wave-particle duality. Rather than assigning probabilities to outcomes, probabilities are instead assigned to entire fine-grained histories. The formulation is fully relativistic and applicable to multi-particle systems. It shall be shown that this new formulation makes the same experimental predictions as quantum field theory, but without having to rely upon the notion of a system evolving in a superposition of quantum states until collapsed by an observation. It is thus free from the problem of deciding what exactly constitutes an observation (the measurement problem) and may therefore be applied just as readily to the macroscopic world as to the microscopic. 
\end{abstract}

\section{Introduction}

  One of the central problems with current quantum theory is the division between the microscopic and macroscopic worlds: at the microscopic level the state of a system evolves according to appropriate wave equations, with states superposing and interfering with each other, whereas at the macroscopic level no such superposition is seen. This division is commonly attributed to the collapse of a system into a definite state whenever it is observed, not that there is any precise definition of what constitutes an observation, and it is thus impossible to say under exactly what circumstances the quantum state will collapse. The final superposition of the states of a quantum system at the time of observation is used to calculate the probabilities of various outcomes. Thus quantum systems are sometimes in a superposition of states, interfering with each other and exhibiting wave like behaviour, and are sometimes in definite states, showing particle like behaviour. This situation is commonly referred to as the wave-particle duality. Such a division is unsatisfactory in its own right, but the aforementioned inability of this picture to tell us when a system is behaving like a collection of waves, and when it is behaving like a collection of particles (when it is being ``observed'') renders the theory incomplete and ambiguous, and so a better formulation of quantum mechanics is required.

  We shall base our formulation on the Feynman path integral formulation of quantum mechanics \cite{bib:feynman}, which essentially states that a set of particles takes all possible paths, which may include the creation and annihilation of particles, and that each possible path has an amplitude associated with it. To find the probability of a particular final distribution of particles, simply add the amplitudes of all the paths that result in that distribution and then square the final amplitude to get the probability. Of course this formulation relies on the notion of a particular particle distribution being selected whenever an observation occurs, with the quantum state consisting of a superposition of distributions at all other times. However would not this unnatural division be eliminated if there was simply a definite distribution of particles at all times? This is clearly the case at the time of observation, perhaps it is also true at every other time, and thus perhaps a particular history is selected rather than a particular final outcome. If appropriate rules were used for calculating the probability of a particular history being selected, they could be compatible with the usual rules for calculating the probabilities for final states being selected.

  It is true that formulations based upon assigning probabilities to histories have been proposed before, most notably the consistent histories approaches of Griffiths, Omn\`es, Hartle and Gell-Mann \cite{bib:griffiths} \cite{bib:omnes} \cite{bib:gellmann}. However all these approaches assign probabilities only to coarse-grained histories, i.e. to the histories of a sequence of events in the macroscopic world. This approach however requires one to draw a line between the microscopic and macroscopic world and this introduces vagueness back into the picture. After all where does one draw the line at which to coarse-grain? Although rules have been developed, they are ones which work for all practical purposes (Bell's FAPP \cite{bib:bell}) only, and are not completely rigorous. Furthermore one is still left with a division between the microscopic and macroscopic worlkds with no clear, single view of reality. Clearly all these difficulties would be removed if one simply applied probabilities to fine-grained histories. It is the claim of consistent historians that this can not be done. However it shall be demonstrated in this paper that there is a logically sound way of assigning probabilities to fine-grained histories which is consistent with the conventional quantum probability law.

  This theory clearly requires that we formulate rules that tell us the probability of a particular history occurring, rather than rules that tell us only about the probabilities for final outcomes. Using the rules given in
Sec. (\ref{new-formulation}), which take into account interference effects, we end up with a definite distribution of particles at all times, with the appearance of wavelike interference effects due merely to the details of the probability rules for selecting particular histories. We then apply these rules to the entire universe over all time, thus choosing a unique cosmic history. The notion of the collapse of the quantum state by observation is thus eliminated. Nor does this formulation, with particles in definite positions at all times, violate the uncertainty principle, for that only concerns what macroscopic observers can know about the state of a system, and as we shall show, the usual quantum rules follow from our new formulation. We shall now derive the necessary rules for calculating the probability for selecting a particular history, and then demonstrate with a few examples how they do indeed give rise to the appearance of wave-like behaviour at the microscopic level.

\section{The new formulation} \label{new-formulation}

  If we are to have rules giving the probability for an entire history rather than for a final outcome, these rules must still give the correct probabilities for final outcomes. Under our new formulation the probability for a final outcome must be the sum of the probabilities of all the histories resulting in that outcome. We can use this fact to derive the probability rules for selecting a particular history. We shall consider space and time cut up into small volume elements of size $\delta V=\delta x \,\delta y \,\delta z \,\delta t$, and then take the limit as $\delta V \to 0$. The whole history is thus cut up into a succession of time slices of duration $\delta t$. At each time slice there are a infinite number of possible particle distributions, and there are an infinite number of ways of picking a set of these distributions which span all time. Each such set constitutes a particular history. In the following analysis we shall for the sake of simplicity ignore any internal polarisations or states of the particles, though it is a mathematically trivial task to extend the method to include these.
  
  In our analysis we shall use the following notation:
  \bigskip 

\indent \makebox[8mm][l]{$\psi( X ) $}\hspace{.5em}=\hspace{.5em}Probability amplitude for $X$ to happen.

  \makebox[8mm][l]{$P( X ) $}\hspace{.5em}=\hspace{.5em}Probability for $X$ to happen. 

  \makebox[8mm][l]{$d_i[t]$}\hspace{.5em}=\hspace{.5em}Distribution $i$, a particular distribution of particles at time $t$. We assume that the possible distributions can each be uniquely numbered such that by summing from $d_1$ to $d_{\infty}$ we include all possible distributions once and once only.

  \makebox[8mm][l]{$ H_i$}\hspace{.5em}=\hspace{.5em}History $i$, where a history is a definite set of particle distributions at each time. We assume that the possible histories can each be uniquely numbered such that by summing from $H_1$ to $H_{\infty}$ we include all possible histories once and once only. 

  \makebox[8mm][l]{$F(H)$}\hspace{.5em}=\hspace{.5em}The standard Feynman amplitude for the history $H$.

  \makebox[8mm][l]{$d_f$}\hspace{.5em}=\hspace{.5em}The final distribution of particles at time $t_f$. We shall calculate the probability of $d_f$ occurring.
\bigskip

Under the Feynman formulation, the probability of a final distribution occurring at $t_f$ is obtained by adding the amplitudes for all the histories resulting in that outcome, and then squaring. This may be expressed as

  \begin{equation} \label{eq:feynmanrule}
   P\left(d_f \right) = \left|\psi \left( d_f \right) \right|^2
    =\left| \sum_H F \left( H \right) \right|^2 .
  \end{equation}

Now instead of summing over all histories at $t_f$, we may obtain the final amplitude by considering all possible distributions at a time $\delta t$ earlier. We can then get the final amplitude by considering for each such distribution the amplitude for it to occur multiplied by the amplitude to get from it to the final distribution. Summing all these amplitudes gives the same final amplitude as above; expressed mathematically this gives

\begin{eqnarray}
   \psi \left( d_f \right)&=& 
     \sum_{n=1}^{\infty} \psi \left( d_n \left[t_f-\delta t\right] \right)
     \psi \left(d_n\left[t_f-\delta t\right]\to d_f\left[t_f \right]\right) \\
   \Rightarrow P\left(d_f \right) &=&
     \left| \psi \left( d_f \right)\right|^2= 
     \left|
     \sum_{n=1}^{\infty} \psi \left( d_n \left[t_f-\delta t\right] \right)
     \psi \left(d_n\left[t_f-\delta t\right] \to d_f\left[t_f\right]\right)
     \right|^2 \nonumber \\
   &=& \sum_{n=1}^{\infty} \Big| 
     \psi \left( d_n \left[t_f-\delta t\right] \right)  
     \psi \left(d_n\left[t_f-\delta t\right] \to d_f\left[t_f\right] \right)
     \Big|^2 I\left( d_f \left[t_f \right]\right) \nonumber \\
     \label{probs} 
  &=& \sum_{n=1}^{\infty}  
     P \left( d_n \left[t_f-\delta t\right] \right)  
     P \left(d_n\left[t_f-\delta t\right] \to d_f\left[t_f\right] \right)
     I\left( d_f \left[t_f\right]\right) .
 \end{eqnarray}
All amplitudes in the above and following are to be calculated using the usual Feynman rules, with the amplitude for a distribution at a particular time obtained by summing over histories only up to that time. In the above $\psi(A \to B )$ is the amplitude for $A$ to evolve into $B$ over the time interval $\delta t$, and $I$ is simply

 \begin{equation} \label{eq:I}
   I \left( d_i \left[t \right] \right) \equiv 
     \frac{ \left|
     \sum\limits_{n=1}^{\infty} \psi \left( d_n \left[t-\delta t\right]\right)  
     \psi \left(d_n\left[t-\delta t\right] \to d_i\left[t\right] \right)
     \right|^2}
     {\sum\limits_{n=1}^{\infty} \Big| 
     \psi \left( d_n \left[t-\delta t\right] \right)  
     \psi \left(d_n\left[t-\delta t\right] \to d_i\left[ t \right] \right)
     \Big|^2} .
 \end{equation}
We shall call $I$ the interference factor. It is different for every possible distribution and is different at different times. It is a measure of how much interference between the different possible histories that contain the distribution of interest there is at each time. With Eq.\ (\ref{probs}) we have expressed the probability of a final distribution $d_f$ in terms of the probabilities of the distributions $\delta t$ earlier. We may apply the same logic to these probabilities, expressing them in terms of probabilities of distributions another $\delta t$ earlier, and so on all the way to our boundary conditions at $t=0$. We should note here that when we refer to probabilities of distributions we are referring to the probability as calculated by standard Feynman rules, however when we obtain our result for probabilities for histories it will emerge that the actual probability for a distribution at an intermediate time will in general only coincide with the Feynman probability when a measurement is being made on the distribution, thus the quantities in Eq. (\ref{eq:I}) are not necessarily true probabilities, but this does not matter as long as the rules for calculating them are well defined, as indeed they are. This will require us to sum over all possible distributions at each time slice. We shall number the time slices from $0$ to $T$, with the time at time slice $j$ being $j\, \delta t$, thus $T=t_f / \delta t$. The summation index for the distributions at time slice $j$ shall be $n_j$. The boundary condition may consist of a definite distribution, or a superposition of distributions, so we must therefore sum over the number of superposed states at $t=0$, and this number we shall label $\beta$. Thus we have

\begin{equation}
  P \left( d_f\right) = \sum_{n_0 =1}^{\beta}  \sum_{n_1=1}^{\infty} 
   \ldots \sum_{n_{T-1}=1}^{\infty}
   P\left( d_{n_0} \right) \prod_{\alpha=1}^T 
   P\left( d_{n_{\alpha-1}} \to d_{n_{\alpha}} \right)
   I\left(d_{n_{\alpha}} \right).
\end{equation}
There is of course no summation over $d_{n_T}=d_f$, but it still appears in the product of probabilities. As we can see, we have now expressed the probability of a final outcome as a sum of probabilities. We sum over every possible intermediate distribution, i.e. over every history. Thus this is equivalent to summing the probabilities for each history, with the probability of a history given by

\begin{eqnarray}
  P \left( H\right) &=& 
  P\left( d_{H,0} \right)
  \prod_{\alpha=1}^T P\left( d_{H,\alpha-1} \to d_{H,\alpha} \right)
  I\left(d_{H,\alpha} \right)  \\ \label{history-probability}
  &=& \left| F \left( H \right) \right|^2 \prod_{\alpha=1}^T 
   I\left(d_{H,\alpha} \right),
\end{eqnarray}
where $d_{H,\alpha}$ is the distribution at time slice
$\alpha$ in history $H$. This result is the usual Feynman amplitude squared times the product of all the interference factors.

\subsection{Is the probability law well-defined and consistent with observation?}

We have succeeded in Eq. (\ref{history-probability}) in assigning probabilities to fine grained histories in a manner which recovers the usual Feynman probabilities at the final time $T$. However one might legitimately ask how this formalism accounts for probabilities at intermediate times. After all, we have advocated taking the limit of the final time going to infinity --- not a condition realised in very many actual experiments. Also one might wonder if the method is well defined in the limit $\delta V \to 0$, and in the limit $T \to \infty$.

The question of probabilities at intermediate times shall be considered in depth in Sec. (\ref{self-interference}). The basic idea is that in any quantum measurement situation there is no significant future interference between the various possible histories containing the various possible outcomes of the measurement (the decoherence effect), and thus one need not consider what happens in the future after performing the experiment. Thus the results are the same (with a negligible error factor) as they would be if the measurement was made at the end time $T$.

The question as to whether the history probabilities are well defined in the limit $\delta V \to 0$ is closely related to the question of whether the Feynman path integral approach itself is well defined in this limit. As we can see, Eq. (\ref{eq:feynmanrule}) is just an expression of the usual Feynman rule, and Eq. (\ref{history-probability}) is derived simply by a rearrangement of terms, to express Eq. (\ref{eq:feynmanrule}) as a sum of real numbers rather than as the square of a sum of complex numbers. Thus as long as Eq. (\ref{eq:feynmanrule}) is well defined, so should Eq. (\ref{history-probability}) be.

To be more rigorous, let us consider a small degree of coarse graining. After all to assign a probability to a single fine-grained history in the limit of $\delta V \to 0$ is a bit odd as the probability will go to zero. Rather than assigning probabilities to individual fine grained histories we can assign probabilities to families of histories defined by particle positions to within a specified range $x \to x + \delta x$. Also consider the times to be defined to within the range $t \to t + \delta t$. The probabilities of these coarse grained histories will be the sums of the probabilities of the fine grained histories which are compatible with them, i.e. those in which the particle positions fall within the specified ranges. If we have coarse grained histories of accuracy $\delta x$, then we can calculate final outcome probabilities for particle positions within the range $x \to x + \delta x$. Now let us set the size of the coarse graining to be well below the limits of any experimental detectability, for example let $\delta x = 10^{-20}$ m. To extract any experimental prediction with an error of say $10^{-4}$ m we would of course integrate over a large number of these coarse grained final outcomes. Clearly taking the limit $\delta V \to 0$ won't affect these results as we can use the original Feynman approach which is known to be well defined. This thus means (due to the transformation) that the probabilities for all the coarse grained histories are well defined. Yet we can make the coarse graining so small that it is effectively fine grained. No matter how small (it could be $10^{-1000}$ m)  it is still well defined, so the probability rule is well defined in the limit $\delta V \to 0$.

Finally, what of the limit $T \to \infty$? The situation here is less clear. It is certainly plausible that the limit is well defined, as at a large cosmological time, in an expanding universe, we could imagine all particles being so far separated that they no longer interfere with each other (or even themselves) to any appreciable degree. We can see that if ever a point is reached at which future interference factors can be ignored, or are in fact all zero, then we can calculate the probabilities for histories up to that time in a well defined way. In the limit of $T \to \infty$ this ideal might be reached. However such considerations do depend on a choice of cosmological model, and given the current uncertain state of cosmological research we cannot draw any definite conclusions. Nevertheless, it remains the case that even if the limit of $T \to \infty$ is not well defined, and we thus have to introduce an arbitrary parameter $T$, it will be an arbitrary parameter of no practical consequence. This is for the same reasons as alluded to above concerning the correctness of the probability law for intermediate times. In any quantum measurement situation any future interference factors between the observable outcomes are effectively zero. Thus it makes no practical difference where we make our $T$ cutoff point: all experimental observations will still produce the same results.

\subsection{Quantisation of Second Order Differential Equations}

The result of Eq.\ (\ref{history-probability}) is based upon a path integral formulation in which the amplitude for each distribution to distribution step is dependent only on the two distributions concerned, and independent of what happened previously. Such a formulation would follow from the quantisation of a first order differential equation such as the Schr\"odinger equation, but the quantisation of a second order differential equation such as the Klein-Gordon equation would result in different behaviour. Particles would now effectively have a velocity as well as a position at each time, and both of these would determine the probability distribution for the next time slice. The velocity would be determined by where the particles were in the previous time slice, and thus the probability of moving from a distribution at time $t$ to one at $t+ \delta t$ would depend on these two distributions and on the distribution at $t- \delta t$. Thus if we consider an entire history $H$, its Feynman amplitude may be decomposed into the product of amplitudes for each step as before, but now with the amplitude law of the form

\begin{eqnarray}
  && \psi \left(d_H \left[t \right] \to d_H \left[ t + \delta t \right]
    \right)=
    f \left( d_H \left[ t - \delta t \right], d_H \left[ t \right],
    d_H \left[ t + \delta t \right] \right) \nonumber \\ 
  &&  \equiv
    \psi \left(d_H \left[t \right] \to d_H \left[ t + \delta t \right] \right)
    \left( d_H \left[ t - \delta t \right], d_H \left[ t \right],
    d_H \left[ t + \delta t \right] \right),   
\end{eqnarray}
where we have explicitly indicated the dependency on the immediately preceding distribution in the notation for the amplitude. We may now repeat our previous analysis but we must consider a final pair of distributions at $t_f$ and $t_f -\delta t$, called $d_f$ and $d_{f-1}$, instead of a single final distribution. The amplitude for a final outcome which has this pair of final distributions is

\begin{equation}
 \psi \left( d_{f-1},d_f \right) = \sum_{n=1}^{\infty} 
  \psi \left( d_n \left[ t_f -2 \delta t \right],
  d_{f-1} \left[t-\delta t \right] \right)
  \psi \left(d_{f-1} \left[ t-\delta t \right] \to d_f \left[t \right]\right)
  \left( d_n,d_{f-1},d_f \right),
\end{equation}
where $\psi (d_i,d_j)$ is the amplitude for the distribution pair $d_i,d_j$ to occur at the specified pair of adjacent time slices. As before we square this to get the probability:

\begin{eqnarray}
 P\left(d_{f-1},d_f \right) &=& \left| \psi \left(d_{f-1},d_f \right)\right|^2
  =\left| \sum_{n=1}^{\infty} 
  \psi \left( d_n ,d_{f-1} \right)
  \psi \left(d_{f-1} \to d_f \right)
  \left( d_n,d_{f-1},d_f \right) \right|^2 \nonumber \\
 &=& \sum_{n=1}^{\infty} \Big| 
  \psi \left( d_n , d_{f-1}\right)
  \psi \left( d_{f-1} \to d_f \right)
  \left( d_n,d_{f-1},d_f \right) \Big|^2 
  I \left(d_{f-1},d_f \right) \nonumber \\
&=& \sum_{n=1}^{\infty} 
  P \left( d_n , d_{f-1}\right)
  P \left(d_{f-1} \to d_f \right)
  \left( d_n,d_{f-1},d_f \right) I \left(d_{f-1},d_f \right),
\end{eqnarray}
where for simplicity we have not explicitly specified the times, and $I$ is the usual interference factor and is given by

\begin{equation}
I \left(d_i \left[ t-\delta t \right], d_j \left[t \right] \right) \equiv
\frac{ \left| \sum\limits_{n=1}^{\infty} \psi 
 \left( d_n \left[ t-2 \delta t \right], d_i \left[ t-\delta t \right]
 \right) \psi \left(d_i \to d_j \right)
\left( d_n,d_i,d_j \right) \right|^2}
{\sum\limits_{n=1}^{\infty} \Big| \psi 
\left(d_n \left[ t-2 \delta t \right], d_i \left[ t-\delta t \right] \right) \psi \left(d_i \to d_j \right)
\left( d_n,d_i,d_j \right) \Big|^2}.
\end{equation}
The interference factor now applies to a pair of distributions in adjacent time slices, whereas in the first order differential equation case it applied only to a single distribution.

We have now succeeded in expressing the probability of obtaining a pair of distributions at $t_f$ and $t_f - \delta t$ in terms of the probabilities of pairs $\delta t$ earlier. As before we may apply this logic recursively, and again we will obtain the product of the probabilities of the individual steps in the path, which gives the usual Feynman amplitude squared, multiplied by the product of the interference factors, where these are now defined on a pair of distributions rather than on a single distribution.

We can of course extend this analysis to the quantisation of differential equations of any order. Thus we have a formulation for calculating the probabilities for entire histories which is applicable to any quantum field theory. Naturally we cannot perform these calculations without knowing the boundary conditions: these are the initial state of the universe, which may be a definite distribution of particles or a superposition of possible distributions. Of course we don't know what the cosmic boundary conditions are, but ignorance of them does not prevent us from making effective use of this theory, as when conducting a quantum experiment we can simply ask: given that the universe has evolved to the point of the start of the experiment, what then are the probabilities of the various possible outcomes? To answer this question we need only consider the interference factors and Feynman amplitudes over the course of the experiment, subject to certain provisos that will be discussed in Sec. (\ref{self-interference}). 

\section{The physical significance of the interference factors}

\begin{figure}
 \includegraphics[0,0][10cm,5cm]{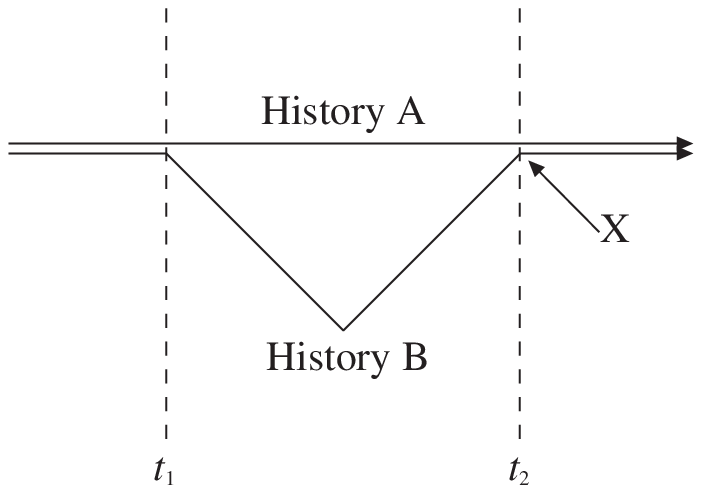}%
 \caption{ \label{two-histories}
  Two interfering histories.}
\end{figure}

\begin{figure}
 \includegraphics[0,0cm][10cm,5cm]{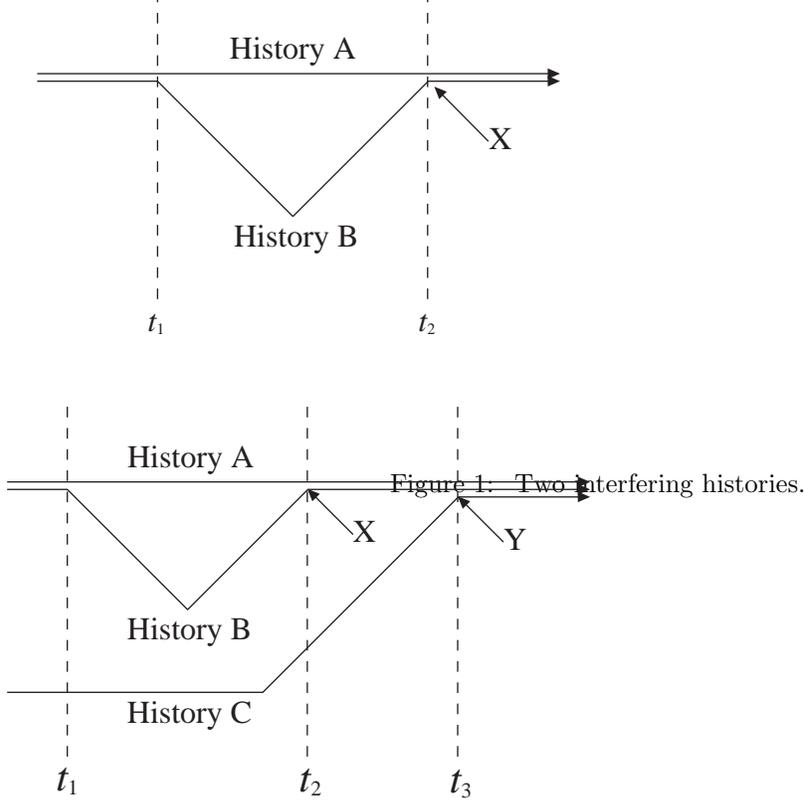}%
 \caption{ \label{three-histories}
  Three interfering histories.}
\end{figure}

We can see that the inclusion of the interference factors is mathematically correct, and it will now be instructive to examine the physical consequences of them. When selecting a history, we can not of course select one which ends in a final distribution of zero probability due to interference with other histories at this time, but equally, we can not allow such a thing to occur at any point in the history. To illustrate some of the physics of the interference factors, we shall consider some simple examples involving the motion of a single particle which can interfere with itself. Consider Fig. (\ref{two-histories}): it represents two histories, which describe the motion of the particle, which diverge at time $t_1$ and converge again at point $X$ at time $t_2$ with a phase difference of $\pi$. They are thus exactly out of phase and cancel, and there are no other histories with any significant amplitude to be at $X$ at time $t_2$. If we calculate the interference factor at $X$ and $t_2$, it is

\begin{equation}
I=\frac{\left( \frac{1}{\sqrt{2}}-\frac{1}{\sqrt{2}} \right)^2}
 {\left(\frac{1}{\sqrt{2}}\right)^2+\left(\frac{-1}{\sqrt{2}}
\right)^2 } =0.
\end{equation}
Thus the probability of picking either of these histories is zero, because due to destructive interference there is no possibility of the particle passing through the point $X$ at time $t_2$.

Could we not simply consider the interference of histories at the end time, rather than keeping track of it at every time? In the above example this would indeed give the same result, but in general we cannot because other histories could subsequently interfere, as shown in Fig. (\ref{three-histories}). Here we have added a third history to the previous diagram, with the same phase as history $A$. It converges with the previous two at $Y$ after their interference event at $X$. Neither history $A$ nor $B$ should have any amplitude to occur as they destructively interfere at $X$, but if we now look at history $A$, and consider interference between entire histories at time $t_3$ the interference factor is

\begin{equation}
I=\frac{\left(\frac{1}{\sqrt{3}}+\frac{1}{\sqrt{3}}-\frac{1}{\sqrt{3}}
 \right)^2}
 {\left(\frac{1}{\sqrt{3}}\right)^2+\left(\frac{1}{\sqrt{3}}\right)^2
 +\left(\frac{-1}{\sqrt{3}}\right)^2}=\frac{1}{3}. 
\end{equation}
This would give history $A$ a non-zero probability of occurring. This is why we must take account of interference at the point it occurs: if we do so at a later time we have no idea where it happened and thus which histories should be favoured or penalised by the interference.
   
When calculating the interference factor for a particular distribution $d(t)$ of particles at time $t$, we consider all possible distributions at a time $\delta t$ earlier and multiply their amplitudes by the amplitude to get from them to the state at $t$. These amplitudes are then used to calculate the interference factor, but what would be wrong with considering the interference between all histories that result in $d(t)$ at time $t$, rather than just considering what happens in the time interval $t-\delta t \to t$? This method would in fact result in overcounting of interference factors. For example consider Fig. (\ref{two-histories}) again, and imagine that this time the two histories do not completely cancel out, but that the resultant amplitude at $X$ is half what it was before the paths split. This results in an interference factor of $1/4$. Now consider what happens if we calculate the interference factor between the two histories a little time after $t_2$: it is again $1/4$, and is indeed $1/4$ at every time beyond $t_2$. Clearly if we multiply all these factors together we get something very close to zero, which is not the result we want. So the interference factor gets counted many times if we calculate it over whole histories rather than over small time increments. Intuitively we deal with small time increments because we only want to take account of the interference at the time it occurs --- we don't want to count it several times.

\begin{figure}
 \includegraphics[0,2cm][10cm,8cm]{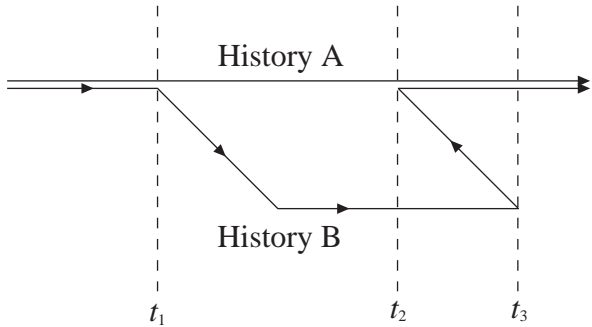}%
 \caption{ \label{backward-history}
  A particle which appears to go back in time.}
\end{figure}

In the Feynman diagram formulation of quantum mechanics, which will of course give the same amplitudes as considering every path, it is common to consider diagrams where particles appear to travel back in time and then interfere with each other; an example of this is shown in Fig. (\ref{backward-history}). Our formulation however only deals with interference factors due to influences from the past, so does it predict the correct behaviour in situations like this? In fact it does: in Fig. (\ref{backward-history}) there is no interference factor at $t_2$ because at that point history $A$ has a single particle whereas history $B$ instantaneously has two particles present, and we can only have interference between histories with the same distribution of particles, however there is an interference factor at $t_3$, as here history $B$ goes from containing three particles, to instantaneously two, and then to one in the same position as history $A$'s single particle. With two different histories evolving into the same distribution we get an interference factor, which is clearly going to have the same effect as the usual Feynman rules in this case. Of course this example presents no conceptual difficulty if we view history $B$ as consisting of a single particle up to $t_2$, at which point a particle-antiparticle pair is produced, the antiparticle of which then annihilates the original particle at $t_3$, leaving us with the surviving particle of the pair. From this point of view there is no need to consider anything to be going back in time, and this is how such situations should be viewed in general for the purpose of calculating interference factors.

\section{A new paradigm for quantum mechanics}

This formulation of quantum mechanics gives precise rules for calculating the probability of an entire cosmic history; it applies to the microscopic and macroscopic worlds equally well and there is no longer any ambiguity about when exactly a collapse of the quantum state occurs --- indeed this concept no longer exists. It is thus axiomatic and, as mentioned, does not rely on the presence of any observer or experimental apparatus to collapse the quantum state. If this is so, why is it that the quantum state appears to collapse when observed in situations such as the famous two-slit experiment? We shall apply the above formulation to a variant of the two-slit experiment in a moment to demonstrate that it works, but first it is instructive to consider some of the wider implications of this theory.

Firstly waves have been entirely dispensed with: in the universe in which we live there are simply particles moving around the place. Even in the two-slit experiment with a single photon, the photon is behaving as a particle all the time, going through one slit or another, eventually arriving at a detector where it is observed. There has been no collapse of any wave, what we see when we observe the photon, i.e. a particle, is what it always was: there never was a wave, and there is no mysterious collapse of the wave going on when the photon reaches the detector. We are fooled into thinking that there must have been a wave before the photon reached the detector because if we repeat the experiment many times the distribution of the photons' positions at the detector is that which we would expect from a wave, but this interference pattern arises as a result of the universe picking an entire path or history, of which the two-slit experiment is a small part, not as the result of a wave locally interfering with itself and then collapsing to a particle. This formulation of quantum mechanics states that at all times there are only particles, and indeed all quantum experiments \emph{always} observe particles --- nobody has ever seen a quantum mechanical wave. However this theory is definitely not an ordinary particle theory: there are no equations of motion for the particles, their distributions are not determined by local laws. The local evolution of a group of particles will be most likely to happen in such a way as to maximise the probability of the entire history of which that motion is a part, but this is something which can only be determined by considering the whole history at once, so the local evolution of particles is determined by non-local factors. There is also of course the usual randomness associated with quantum mechanics in the evolution of particle distributions. There is also no concept of time evolution, the state of the whole universe over all time and space is set all in one go, the appearance of the arrow of time is therefore a consequence of the boundary conditions at zero time, whatever they may be, favouring the selection of histories in which there appears to be a flow of time. Also the particles are always in well defined positions, a definite history has been chosen. This is after all what we observe with our own eyes, we never see any sign of quantum states being superposed at the macroscopic level, and as far as we can tell the universe is following a particular course, appearing to superpose quantum states only at the microscopic level. Thus this formulation is consistent with our
macroscopic observations, and we shall now show in detail how it gives rise to the appearance of wave-like behaviour at the microscopic level.
 
\section{A photon interfering with itself} \label{self-interference}

\begin{figure}
 \includegraphics[0,14cm][5cm,19cm]{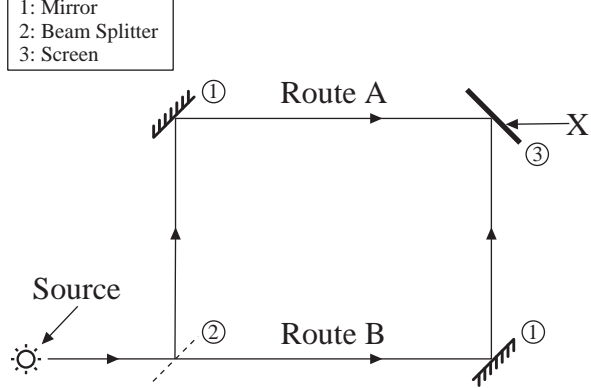}%
 \caption{ \label{interference-experiment}
  A version of the two-slit experiment.} 
\end{figure}

To illustrate the application of this theory to real experiments we shall consider a variation on the two slit experiment in which an atom emits a single photon, which meets a beam splitter, and can from there reach a detecting screen by one of two routes where an interference pattern will form, as shown in Fig. (\ref{interference-experiment}). We have labelled the two possible routes as route $A$ and route $B$. We shall consider the probability for the photon to arrive at a fixed point on the screen marked $X$. We also assume that all the parts of the apparatus are connected, perhaps by being bolted to a common floor.

It is of course hardly convenient to analyse this problem from a pure path integral approach, considering every possible path of every particle involved in the experiment. Fortunately there are simplifications that can be made. Firstly we can use Feynman diagrams rather than path integrals to perform our calculations, our theory requires us to know the amplitude for each possible distribution immediately prior to an interference event, and these amplitudes can be calculated with Feynman diagrams. Secondly we do not even need to know all the details of the Feynman diagrams. For example the internal details of the Feynman diagrams within the source atom do not concern us, as it is known that they will result in bound states with phase rotation given by $\psi(t)=\psi(0) \exp (i E t / \hbar )$, where $E$ is the energy of the atom. We may consider the source atom to start in one quantum state, and to then drop down to a lower energy state emitting a photon in the process. We need only know what state the atom is in, and what its phase/amplitude for being in that state is. Also, we do not need to worry about the intricacies of the quantum state of the experimental apparatus: if we can demonstrate that its quantum state is independent of which route the photon takes, then the internal details of that quantum state will be of no concern, a point which we shall discuss in more detail later. Finally we shall use the standard photon propagator to describe the probability distribution of the photon's journey, and we shall denote the time of arrival of the photon at the screen by $t_s$. We shall also denote the time at which the experiment begins by $t_E$.

The approximation procedure used here, whereby we ignore irrelevant microscopic aspects of the experimental setup, bears some similarity to the coarse-graining procedure used in the Gell-Mann and Hartle consistent histories approaches \cite{bib:gellmann}. However there are significant differences. In our formulation the coarse-graining procedure is introduced as a convenient approximation procedure to simplify the analysis of a problem, however there is in principle no need to use it. If one wished then one could analyse a problem using fine-grained histories only. Of course it would be impractical to actually do this for all but the simplest idealised problems, but the difficulties are purely practical ones, not fundamental ones. This stands in contrast to the consistent histories approach which denies the possibility of assigning probabilities to fine-grained histories at all. Also we are not coarse-graining everything here: we are considering the fine-grained history of our photon, whereas in the consistent histories approach one would consider only the coarse-grained histories specifying different detection points on the screen.

\subsection{The phase difference between the two routes}

Considered from this point of view, what gives rise to a phase difference between the two routes a photon can take to a point on the screen? The photon propagator's phase contribution is independent of its flight time, i.e. the phase of a photon is fixed at its moment of creation. The photon propagator also gives a very large probability for the photon to travel at the speed of light, with a negligible contribution for other speeds, so we shall for simplicity simply assume the photon to travel at exactly the speed of light between parts of the apparatus. Thus the phase difference due to the different flight times between two diagrams with the photon going a different route in each is in fact due to the quantum state of the emitting atom. Strictly speaking phase is a property of the whole system, not of individual components of it, but as the events that give rise to the phase difference occur inside the atom in this case, it is useful to consider the phase difference to be carried by the atom. There is also an additional phase difference between the two routes, caused by the effect of the beam splitter on the photon's phase.

A typical atom might radiate at a frequency of $10^{15}$Hz over a period of $10^{-9}$s. Thus there would be the order of $10^6$ wavelengths of light in the light pulse. This is of course the classical picture: what is happening quantum mechanically is that the atom emits the photon at a random time, with a half-life of $10^{-9}$s. Thus if we sum over all possible paths we will be getting interference from histories in which the atom emitted the photon at different times. For example a history in which a photon took route $A$ to point $X$ in the experiment might interfere with one in which it took route $B$. If the photon was emitted at a different time in each of these two histories, then it can only end up at the same place on the screen at the same time in each history if the route it took to get there is of different length in each case, such that the extra journey time for the history where it was emitted first is equal to the time difference $\Delta t$ between the emission in the two cases.
 
Now the atom's wave function $\psi$ evolves in time as $\psi(t)=\psi(0) \exp ( i E t/ \hbar)$. This means that the rate of phase change of the atom will be higher before it emits the photon than after. Thus the final phase of the atom differs by $\Delta E \Delta t / \hbar$ between the two histories, where $\Delta E$ is the energy difference between the two atomic states, which is equal to the energy of the photon. The phase difference caused by different flight times is due to the extra time for which the atom in the history where the photon is emitted later is in a higher energy state. This plus the phase difference caused by the beam splitter gives us the total phase difference between the two routes to point $X$.

\subsection{Apparatus is unaffected by the route of the photon}

So far we have demonstrated that the quantum state of the atom is the same whether the photon took route $A$ or $B$ to reach point $X$, apart from a phase shift, however there will not be an interference factor when these two histories reconverge at the detecting screen unless everything else is the same whichever way the photon goes: if the photon going route $A$ caused some disturbance that taking route $B$ wouldn't then the universe would evolve differently from thereon and we would end up with a different set of particle distributions at the time of arrival at the screen depending on which way the photon went, which would mean no interference. 
  
  We shall consider the effect on the apparatus that the photon bouncing off the mirrors and beam splitters might cause. When it bounces off a mirror or beam splitter it imparts some momentum to the apparatus. It may bounce a few times in the above diagram before reaching the detector, but when it does the whole apparatus has received a net momentum equal to the photon's momentum. We are also assuming here that the detecting screen has no way of knowing which way the photon went to reach it, and that the photon excites the same electronic states in the absorbing material either way. If this is not the case, the apparatus can always be rearranged such that the paths arrive at the screen from the same direction. Thus the overall momentum and energy of the entire apparatus will be the same either way and the detector will be in the same superposition of states after absorbing the photon. The only question remaining is: could the photon bouncing off a mirror and imparting momentum to it set up vibrational states in the apparatus? Then we might expect different vibrational states to be set up depending on which way it went. In fact this does not happen because in bouncing off a mirror the photon does not impart enough energy to excite the lowest energy vibrational state of the apparatus. Let us show this for a typical example. Imagine our photon has a wavelength of 400nm, its wavevector $k$ is given by

\begin{equation}
  k=\frac{2 \pi}{\lambda}= \mathrm{1.57 \times 10^7 m^{-1}},  
\end{equation}
and its momentum is given by
\begin{equation}
  p=\hbar k=\mathrm{1.66 \times 10^{-27} \,kg\,m\,s^{-1}}, 
\end{equation}
If it bounces off a mirror at right angles it imparts $\sqrt{2} p$ of momentum to the apparatus. A lightweight apparatus, free to move in space might weigh 100g (Of course a real apparatus would probably be firmly attached to the Earth and would have an effective mass equal to that of the Earth, but we want to illustrate the point with a set-up where the effect is greatest, and show that it is still negligible), if so it will pick up a velocity of
\begin{equation}
  v=\frac{\sqrt{2}p}{m}=\mathrm{2.34 \times 10^{-26} \,m\,s^{-1}}.
\end{equation}
From this we can find the energy imparted to the apparatus:
\begin{equation}
  E=\frac{mv^2}{2}=\mathrm{2.75 \times 10^{-53}\,J}. 
\end{equation}
A quantum state with this energy would have an angular frequency of
\begin{equation}
  \omega=\frac{E}{\hbar}=\mathrm{2.60 \times 10^{-19}\,rad\,s^{-1}}.
\end{equation}
Clearly the lowest energy mode of vibration of the apparatus is going to have vastly greater frequency than this, so the energy may go into overall momentum, but not into an excited state of the apparatus. It is due to considerations like this that it is important that all the parts of the apparatus are connected. For example if we conducted this experiment in space and all the mirrors were floating freely some of them would pick up momentum when a photon bounced off them and go drifting off, so the final state of the apparatus would depend on which way the photon went.
 
As mentioned earlier, the detecting screen works in such a way that arrival of the photon at the screen will have the same effect on the screen whichever route the photon took. However if the detector was constructed in such a way that it would be put into a different quantum state depending on which direction a photon entered it from, then we would lose interference.

From this we conclude that the quantum state of the entire apparatus and therefore of the entire universe, as the path the photon took has had no effect on anything else in the cosmos, is exactly the same at $t_s$, apart from a phase shift, whichever way the photon went to reach point $X$. Of course there will in fact be many possible states associated with each route: the apparatus could evolve in many different ways at the microscopic level, but they will be the same set of ways in each case, and there will always be the same interference factors between pairs corresponding to route $A$ or route $B$ being taken.

It is also clear that in this formulation of quantum mechanics only histories which have the macroscopic experimental apparatus evolving classically have any significant probability to occur, those which involve the particles in the apparatus moving in unusual ways will destructively interfere with each other, whereas  those that result in the whole apparatus behaving classically will constructively interfere.

\subsection{Observation and decoherence}

If we could consider only the time over which the experiment took place, without regard for what happens before or after the experiment, we could set boundary conditions at the time the experiment began, and by the analysis of Sec. (\ref{new-formulation}) we would find that we could use the usual Feynman rules to calculate the probability of the photon arriving at $X$, which is the same thing as finding the probability that a history containing the event of the photon arriving at $X$, without regard for what happens at other times, is selected. Thus we would recover the usual quantum probability law. However we really want to consider the whole of the universe at once. This means that the probability for the photon to reach a point on the screen is really the probability for it to reach that point on the screen multiplied by the sum of the probabilities of all the possible future developments thereafter, and it might be that these probabilities depend on which point on the screen the photon arrives at, in which case  histories containing the event of the photon arriving at particular places on the screen might have different probabilities to occur than expected. However in practice this consideration makes no difference.
 
We shall consider the possible futures after the end of the experiment, for the moment imagine that we have a boundary condition for the photon to be at $X$ at $t_s$, i.e. we are ignoring possible future interference effects with the evolution of other states that have an amplitude to occur at $t_s$, and we are ignoring what happened before the photon arrived at the screen. In this scenario the sum of all the probabilities of all the histories starting with the photon at $X$ and evolving to anything at all at infinite time is unity. Now consider the time from $t_E$ to infinity: the histories to be summed are the same from $t_s$ to infinity, with each of these now having the set of ways the photon can go from the emitting atom to $X$ prepended to it. This multiplies the probability of each history, considered from $t_s$ to infinity, by the probability for the photon to go from the emitter to $X$. As was shown, this will just be the probability calculated from the usual Feynman rules.

Now in real life the boundary condition is set at $t_E$, not at the time of detection, so we might expect there to be some interference between the various possible outcomes of the experiment after the experiment has completed. The quantum states of the apparatus for each possible outcome will evolve into a probabilistic superposition of many possible states, with some of these overlapping with the states due to different outcomes. It is at this point in the discussion that the concept of `observation' becomes relevant. For example, imagine that there is some way of measuring where on the screen the photon arrived, and that an experimenter is watching what is going on and writes down this position in a notebook. Clearly the different possible final states of the experiment (considered at the time just after the result has been noted) are going to have negligible amplitudes to evolve into each other, and the possibility of interference between them in the future affecting the probabilities for various final photon positions can be safely ignored. After all, consider two histories, one in which the experimenter records ``the photon arrived 5cm from the left of the screen'' and one in which ``the photon arrived 10cm from the left of the screen'' is recorded. The one history is not likely to evolve into the other: this would involve the spontaneous and organised rearrangement of vast numbers of atoms in the notebook's ink and paper to get from the one state to the other.

So we can see that if we `observe', in the traditional sense of the word, the photon's position, then at that point it will act like a particle, and there will be no further interference from the other possible places the photon could have hit the screen. It looks like the wave has collapsed into a definite state, with all of the photon being localised at a single point.

However it is also clear that there is no need to `observe' in the classical sense of having a conscious entity register the final position of the photon. As long as the different positions the photon can arrive at put the apparatus into different states that have negligible amplitude to evolve into each other, then we can analyse the situation as if the photon was a wave until it hit the screen, thus requiring interference between different paths to be considered, but then collapsed into a particle upon hitting the screen, thus allowing us to forget about the other places the photon might have ended up, as the histories due to these can no longer interfere with each other.

Therefore in the example considered the probability of the photon arriving at point $X$ is just the usual quantum probability law. Also the photon has actually gone one of the two routes, but this is completely undetectable experimentally: we have after all made a point of demonstrating that the state of the rest of the universe is unaffected by which way it went.

The fact that there is negligible probability for the different histories corresponding to the different experimental outcomes to overlap again is called decoherence and it plays a prominent role in several other approaches to quantum mechanics \cite{bib:kampen} \cite{bib:zeh} \cite{bib:zurek}. It is found to occur whenever a particle becomes entangled with its environment, or when frictional or dissipative processes occur. In most formulations of quantum mechanics the role of decoherence is to explain why we do not observe quantum mechanical interference at the macroscopic level. In our formulation it does not play this role, as it is clear from the axioms of the formulation that definite particle positions will always be observed rather than superposed states, instead it plays the role of ensuring that it is safe for us to ignore what happens next when calculating probabilities for outcomes of quantum experiments, and so we can assign probabilities to histories, and thus outcomes, using Eq. (\ref{history-probability}) assuming that the end time is the time of the end of the experiment. Finally it should be noted that although the interference between different experimental outcomes is negligible it is not zero, and so there is a very small error factor in the calculations. We have recovered the usual quantum mechanical rule for assigning probabilities to outcomes of experiments, but we have found it to be only approximately true, though the errors involved are typically extremely small and can thus be ignored.

\subsection{Coarse-graining}

We have also assumed that the experiment takes place, but as we must select an entire cosmic history over all space and time, a history may very well be selected in which the experiment does not take place, so we can of course only ask about the probabilities of various experimental outcomes at $t_s$ given that the experiment begins at $t_E$. We are really asking: what is the probability that a history which has the experiment in it starting at $t_E$ also has the photon arriving at point $X$ at $t_s$? We are not interested in the probability of whether or not the experiment even takes place, only in the possible outcomes if it does. It would be easiest to analyse this if we could consider a particular initial quantum state representing the experiment at $t_E$, ignoring all that has happened earlier, and calculate its probability of evolving into one where the photon ends up at point $X$ at $t_s$. However we can never know the precise quantum state of the experiment at $t_E$, we will only be aware of the macroscopic apparatus, there will be many possible quantum states, differing, for example, in the arrangement of their atoms, that will all correspond to the same macroscopic experimental set up. They will all have different phases and amplitudes to occur depending on their past histories. Thus we will have to consider interference between histories with different quantum states at $t_E$ evolving to the same quantum state at $t_s$ in addition to the already discussed case of a definite state at $t_E$ evolving to a definite state at $t_s$.

We shall analyse this by considering a set of $N$ initial states labelled $S_1, S_2, \ldots ,S_N$. We consider the probability for evolving from these states to a set of states identical in all details except the position where the photon arrives at the screen. We consider this to be a specific final state of the apparatus labelled $S_F$ combined with the final state of the photon, i.e. its position. We label the phase of the photon if it took route $A$ as $\alpha$, and the phase if it took route $B$ as $\beta$. Each initial state of the apparatus will have an amplitude to evolve into the final state $S_F$. We shall denote the initial phase and amplitude of the initial state $j$ by $A_j e^{i a_j}$ and the amplitude for this state to evolve into $S_F$ as $B_j e^{i b_j}$. Thus the total amplitude to evolve from the $j$-th initial state to the state $S_F$ with the photon at a given position is

\begin{equation}
 A_j e^{i a_j} B_j e^{i b_j} \frac{1}{\sqrt{2}} 
   \left( e^{i \alpha} + e^{i \beta} \right)
\end{equation} 

and the total amplitude $\psi$ to get from any state in the collection of initial states to the final state is

\begin{eqnarray}  \label{eq:psi}
 \psi &=& \frac{1}{\sqrt{2}} \left( e^{i \alpha} + e^{i \beta} \right)
 \sum_{j=1}^N A_j e^{i a_j} B_j e^{i b_j} \\  
  &=& \frac{k}{\sqrt{2}} \left( e^{i \alpha} + e^{i \beta} \right)
\end{eqnarray}

Where we have assigned the constant $k$ to the terms under the summation, which are constant and independent of where the photon ends up. Thus we can see that the probability for the photon to hit a particular point on the screen depends only on the phase difference between the two paths to that point and is unaffected by considerations due to the microscopic state of the apparatus. It is also clear from this analysis that the degree of coarse-graining makes no difference to the probability for the photon to hit a particular point on the screen, except insofar as the coarse-grained set of initial states may encompass small variations in the positions of the parts of the apparatus, and small variations in the optical properties of the mirrors and beam splitter.

Of course we should also apply the coarse-graining procedure to the final state of the apparatus as well and consider all the possible fine-grained states corresponding to the classical description of the apparatus. However this simply involves summing Eq. (\ref{eq:psi}) over all of those states, and again the probability depends only on the phase difference between the photon's routes.

\subsection{Relativity}

We have just demonstrated that the usual laws of physics according to the Copenhagen interpretation, are recovered to a very high degree of approximation from our theory. There are only miniscule differences due to the imperfection of the decoherence process. The fact that our theory can be applied to relativistic Feynman diagrams suggests that it is a relativistic theory, however one might question this on the grounds that it is set in a preferred frame. Nevertheless this is not a problem, for if the universe appears relativistic in a particular frame, i.e. all particles obey relativistic dispersion relations, then it is clear that it must also appear relativistic from any other inertial frame.

This argument is not quite rigorous due to the small deviations from the Copenhagen probabilities due to the imperfections of the decoherence process. It is possible that these fluctuations are non-relativistic in nature, and thus relativity might be weakly violated by our theory. However any such violations are clearly well beyond the current limits of experimental detectability.

\subsection{Many particles}

\begin{figure}
 \includegraphics[0,1cm][5cm,6cm]{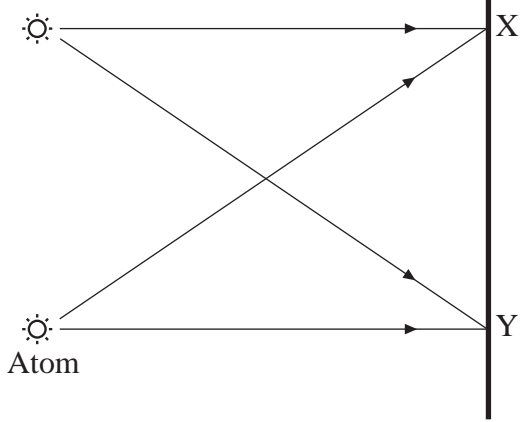}%
 \caption{ \label{two-photons}
  Interference between two photons.} 
\end{figure}

We have just considered the case of a single photon interfering with itself, but what happens when several photons are interfering with each other? As we might expect, the argument is much the same. Consider the apparatus shown in Fig. (\ref{two-photons}). There are two source atoms which are both persuaded to emit a photon at the same time. The two photons emitted will then travel to the screen where there will be interference between them. For example in this case we will consider the amplitude for one photon to hit point $X$ and the other to hit point $Y$. There are two Feynman diagrams which can result in this, and the state of the universe is the same whichever of the two is picked apart from a phase-amplitude factor, so we get the appearance of interference at the screen in the same manner as discussed above, with the precise form of the interference given by the usual Feynman diagram methods. This argument can of course be extended to any number of particles of any type.

\subsection{Many interferences at once}

So far we have considered the case of a single interference experiment, but as we can see from the previous discussion, interference occurs if the rest of the universe is unaffected. The rest of the universe will of course consist of the sum over a vast number of possibilities, but the same possibilities whichever routes our interfering particles are taking. Naturally this sum of cosmic possibilities will include other groups of particles which are also interfering with each other without affecting us or the rest of the universe, thus interference can quite happily happen in lots of different places at once, involving a different set of particles in each case.

\subsection{Interference can be delayed}

We should note that our assumption that the possible futures from each experimental outcome do not coherently interfere with each other, and thus bias for or against histories which contain a particular experimental result, is not always true; indeed one can contrive circumstances where it most definitely is not true. For example imagine that in the experiment just discussed that the length difference between routes $A$ and $B$ was not a few wavelengths but was in fact several metres. In this case the time difference between the flight paths the photon could take would be much greater than the time over which the photon could be emitted, so we would expect there to be no interference pattern at the screen, because there would be negligible probability for a photon taking route $A$ to arrive at the screen at the same time as one taking route $B$. However what if the detector consists of atoms which absorb the photon, and then hold its quantum of energy for a long time, much longer than the time difference between the two routes, and whose interaction with their surrounding environment is unaffected by whether or not they are in this excited state? In this case a photon taking the shorter route would be absorbed by one of these atoms, and would put it into an excited state. A photon taking the longer route would do the same, and the two histories would reconverge once the later photon had been absorbed by an atom, and so you would see an interference pattern at the screen (if you could find out what state the atom had been put into). As we can see, there is no interference of the photon with itself here: when the photon arrives at the screen it will never interfere with another history in which it has reached that same point at the same time by a different route. The interference is in fact caused by what happens next, as in the two possible histories the detecting atom has been in an excited state for a different period of time in each case, thus causing a phase difference.

\section{Which way did the photon go?}

In the above example we considered the case where the route taken by the photon is unknown and interference occurs at the screens, we shall now analyse what happens if the experimenter tries to find out which route the photon took.

\begin{figure}
 \includegraphics[0,0][5cm,5cm]{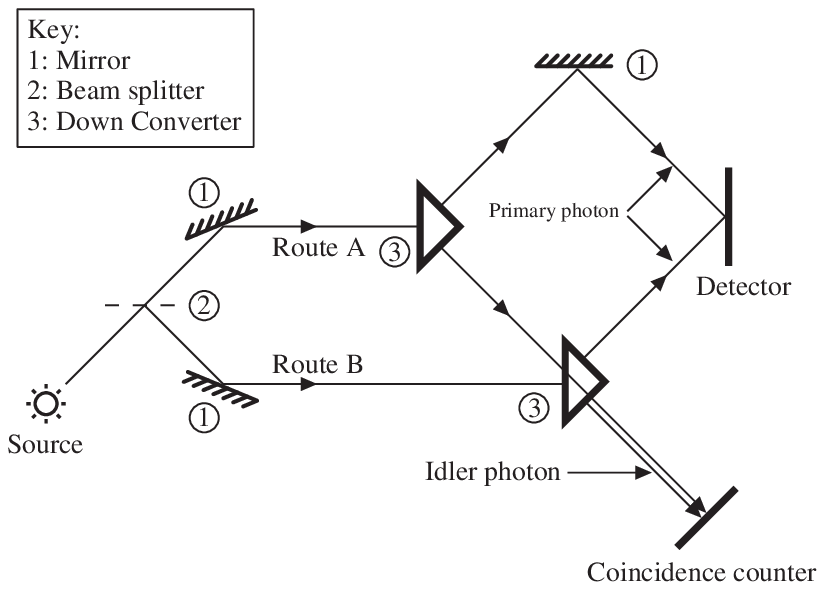}%
 \caption{ \label{idlers-experiment}
  A version of the two-slit experiment with down converters.}
\end{figure}

We shall consider a simple modification of the two slit experiment performed by Mandel et al. In this experiment, pictured in Fig. (\ref{idlers-experiment}), a photon passes through a beam splitter as before, but now each of the routes from here on, labelled $A$ and $B$ as before, takes the photon into a down converter. A pair of photons with half the energy of the original will come out of the down converter. One of the photons out of each down converter goes on to arrive at a detecting screen, this we shall call the primary photon. The other, called an idler photon, is sent to a coincidence counter. We shall denote the time of arrival of the primary photon at the screen by $t_s$. We are assuming in Fig. (\ref{idlers-experiment}) that the detecting screen is put into the same quantum states whichever direction the primary photon arrives from. If both possible idler photons are allowed to reach the coincidence counter then we can't tell which route the original photon took and the apparatus will be in the same quantum state, apart from a phase factor, whichever way it went, so we will get interference as before. In fact this is not strictly true: we must also consider the phase of the photon sent to the coincidence counter. The overall phase of the entire system will depend on the phases of both the primary and idler photons, and we should strictly consider probabilities for pairs of photon positions at the coincidence counter and screen, i.e. the states of the two photons are entangled. However if we make the coincidence counter small enough that across its aperture there is no appreciable phase difference for arriving photons, and only look at the arrival position of a photon on the detecting screen when the coincidence counter is fired, then we can effectively consider in isolation the phase differences between different paths of the primary photon to points at the detecting screen, and will thus observe an interference pattern there as in the standard two-slit experiment.

Now an interesting thing happens if a blocker is put in the path from one of the down converters to the coincidence counter. Let us say route $A$'s idler photon is blocked, then if the original photon takes route $A$, its idler photon will end up in the blocker, if it takes route $B$ its idler photon will end up in the coincidence counter. Thus histories where route $A$ was chosen will not interfere with those where route $B$ was picked, as interference can only occur at the screen if the distribution of all the particles in the universe at $t_s$ is the same, and though both cases may involve a photon arriving at the screen at $t_s$, in one the coincidence counter will have been triggered, in the other it won't have been so there will be no interference. This means that either history $A$ or history $B$ will be chosen, with equal probability if the beam splitter is ideal, but they won't know about each other at the screen, and there will be no interference pattern there, just the usual single slit pattern due to interference between various different ways of taking route $A$ or $B$. In the language of standard quantum theory, the route which the photon went has been observed, so it has collapsed to being either route $A$ or route $B$, but not a superposition of the two.

As we can see from our theory, and have illustrated in some of the above examples, the appearance of the collapse of the quantum state can be caused by microscopic events, unobserved by any conscious entity. If, in the above examples, the taking of one route or another by the original photon results in any difference whatsoever in the final quantum state of the experiment, interference at the screen will be lost. Even if only a single quantum particle is placed differently anywhere in the universe as a result of the photon taking one route rather than another. This difference need not be observed by anybody, and it need not be macroscopic. Most conventional interpretations of quantum mechanics adhere to the notion that the collapse of the quantum state, whether it be a real or apparent phenomenon, is caused either by some direct observation made by a conscious being, or at the very least by events at the microscopic level causing some macroscopic effect. Our formulation is thus in clear disagreement with such theories.

\section{When does quantum mechanical interference occur?}

As we've seen in the preceding examples, sometimes quantum particles will interfere with themselves during the course of an experiment, and sometimes they won't, depending on whether or not anybody tries to find out what they're doing in the meantime. With our new formulation of quantum mechanics we can express this much more precisely:

  ``The standard Feynman rules for calculating quantum mechanical interference amongst a set of particles will be valid if, during the time of interest, that set can reach the same set of destinations by many different routes, and the state of the rest of the universe, apart from an overall phase-amplitude factor, is independent of the routes taken by the particles, provided that there is no subsequent coherent interference between the histories that may develop from the different possible outcomes at the end of the time of interest.''

\section{Non-locality}

This view of quantum mechanics of course will produce the same non-local behaviour as in conventional quantum theory, but as this is such an important feature of quantum mechanics we shall discuss the matter further. Clearly our new formulation of quantum mechanics is non-local: an entire history is selected at once, and thus there are no local equations of motion governing the particles in the universe. To understand what effect this has in practice, it is instructive to consider a particular example. We shall consider a version of the EPR experiment in which a source emits two linearly polarised photons of the same polarisation, and these photons then travel to two widely separated polaroids, which have their polarisation axes aligned parallel to the $x$-axis, where their polarisations are measured. The experiment is set up such that the photons' polarisation will be perpendicular to the $z$-axis, and will thus be in a superposition of polarisation in the $y$ direction, denoted $|y>$ and the $x$ direction, denoted $|x>$. The state of the pair of photons is denoted by $|1>|2>$, where $|1>$ is the polarisation of the first photon, and $|2>$ is the polarisation of the second.

Non-locality is a general feature of Feynman diagrams. In this example there are two possible histories, containing $|x>|x>$ and $|y>|y>$ polarisations of the photons. There are two Feynman diagrams to consider, one where both photons are polarised along the $x$-axis, the other where they are both polarised along the $y$-axis. Now in the usual way we pick a entire cosmic history. There are two sets to choose from, those that have both the photons going through the polaroids, and those that have the photons both being absorbed. We pick one of these histories and we find that the photons either both went through or were both absorbed, there is no possibility of one photon passing through and the other being absorbed. 

It is as if the two measurements have somehow communicated with each other instantaneously to ensure they both produce the same result. For example if we knew that the pair of photons had been produced in the state $(|x>|x>+|y>|y>) / \sqrt{2}$, we cannot then locally look at the state of one of the photons, conclude it is in the state $(|x>+|y>)/ \sqrt{2}$ and thus has half a chance of passing the polaroid, and then also conclude the same about the other photon, and decide that it independently has half a chance of passing its polaroid. This is general feature of Feynman diagrams, you cannot cut them up into pieces and try to consider the behaviour of each piece separately, because states become entangled, as in the current situation where we have the requirement that both photons must have the same polarisation. This sort of behaviour is a consequence of the fact that an entire cosmic history is selected, with the particle distribution fixed everywhere and at all times, and with the probability of a history being chosen depending on the whole. Thus one most certainly can not look at a small section of the history and calculate its probability of occurring, as the probability of an event occurring in a particular place and time could depend on what is happening anywhere else and at any other time in the universe, thus this theory is maximally non-local.

\section{Causality}

In addition to spatial non-locality, we also have non-locality in time: i.e. we have lost causality in our formulation --- the future can affect the past. However the effect of this is confined to the microscopic level. All final experimental results from our formulation must be the same as those predicted by the standard formulation, and as it does not allow causality to be violated, at least from the perspective of a macroscopic observer, then neither does our formulation. We shall now consider this matter in some depth, showing why causality is microscopically violated, and by considering a specific example of this, we shall show the reasons why causality violation must remain confined to the microscopic realm.

In our formulation an entire cosmic history is selected by calculating both the product of probabilities for each step in that history and the product of the
interference factors, which measure interference with other possible histories, at each time. It is the latter factor which makes the theory intrinsically non-causal at the microscopic level. For example a particle, when deciding which branch to take if faced with a choice of going in two directions, which are apparently equally probable from a local perspective, will always choose one route if the other results in a definite destructive interference with another particle at some stage in the future. It is as if the particle ``knows'' what will happen to it in the future if it goes one way or the other. From the perspective of our formulation there is nothing mysterious about this, the probability of a history in which the particle goes one way is zero, the probability of a history in which it goes the other way is non-zero, so at the branching point it always goes one way. However, though this may not be mysterious from a God's eye view of the whole of space-time, it is very mysterious from the local perspective of our particle. From its perspective the probabilities for its various actions now are influenced by what could happen in the future.
 
\begin{figure}
 \includegraphics[0,16cm][10cm,26cm]{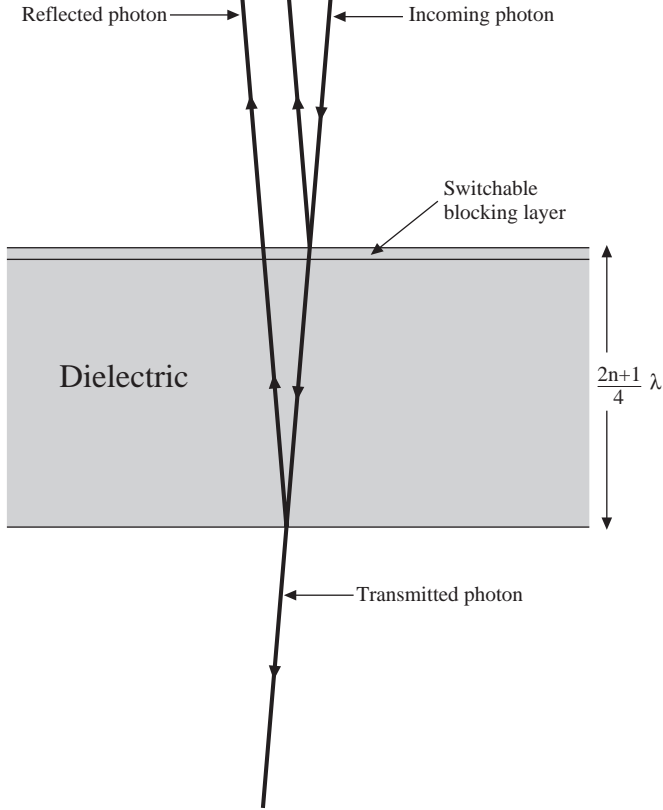}%
 \caption{ \label{fig:dielectric}
  Interference on reflection from a dielectric.} 
\end{figure}

We can consider a particular example of this. See Fig. (\ref{fig:dielectric}), in which one photon at a time is directed into a sheet of dielectric which is $\frac{2n+1}{4} \lambda$ thick, where $\lambda$ is the wavelength of the photon. This arrangement cuts out reflection from the dielectric, as the photon reflecting off the top surface of the sheet will destructively interfere with itself as it can also enter the sheet at the top surface, bounce off the bottom surface and then exit again through the top surface. The histories in which the photon bounces directly off the top and those in which the photon bounces off the bottom surface and then leaves through the top will destructively interfere with each other giving a probability of zero for either eventuality to occur. The interference factor applies when the two histories in which the photon leaves the upper surface merge, its value is zero, so this can't happen, and all the photons will always pass through the lower surface and out the other side of the sheet.

Now if we consider the two possible histories in which the photon is reflected off the sheet, we can see that in the case where the photon enters the sheet and reaches the bottom surface, the probability for it to be reflected there is zero due to the interference which will occur with the other reflective history in the future. At the bottom surface of the sheet the photon always decides to pass through rather than reflect because of something that will happen to it in the future if it reflects.

Imagine a ``Quantum demon'', a being which can observe the properties of quantum particles without in any way disturbing them or any other particles. Such a thing is not possible of course, but hypothetically, if it were possible, then such a demon sitting at the bottom surface of the sheet could watch all the incoming photons, and then by observing how many reflect off the bottom and how many pass through, it would notice that none were being reflected even though the properties of the boundary should allow this. It would thus conclude that no photons are being reflected because if they were they would interfere destructively with other photons or with themselves in other possible histories at some stage in the future. Of course the demon has not actually deduced any more from these observations than it could by simply looking at the entire experimental set up, which would enable it to conclude in advance that no photons will ever be reflected off the bottom surface of the dielectric.

The situation can however be made more interesting if there is a thin layer
of material at the top of the dielectric, which can switched between two states. Either it has the same optical properties as the rest of the dielectric, or it completely absorbs the photons. The state it is in can be set by the experimenter. Imagine that it is switched between the two states very rapidly and randomly. Sometimes a photon will arrive at the top of the sheet when this blocking layer is off, it may then pass through to the bottom surface. At the bottom it may either reflect upwards again or pass through the bottom surface. If it reflects, then if by the time it has reached the top of the sheet again the blocking layer has been switched on again, it will be absorbed by the blocking layer and will not destructively interfere with a later version itself at the top. In this case the photon may reflect from the bottom surface of the sheet and end up being absorbed. On the other hand, if by the time it reaches the top again the blocking layer is still switched off, it will interfere destructively with itself and so it may not reflect off the bottom surface in this case. The interesting point here is that the state the blocking layer is in at the time the photon reaches the top again is not yet set when the photon is at the bottom, yet if it will be on when the photon reaches it, the photon may reflect off the bottom. A quantum demon sitting at the bottom of the sheet would know every time it saw a photon reflected off the bottom that the blocker will be switched on a certain time in the future. In other words a quantum demon could actually see into the future, making deductions about the future from events in the present.

Unfortunately if you or I tried to be a quantum demon and observe whether or not photons were being reflected from the bottom surface of the sheet it wouldn't work. Interference between various possible histories can only occur if all the particles in the universe are in the same positions at the time of interest in each of the histories concerned. When no one is observing, there are two possible histories in our experiment, they diverge when the photon is created (it must be created at slightly different times in each case for it to be in the same place at the same time in the two histories at the end), and merge again when the photon leaves the top of the sheet, either directly from the top, or after having passed down through the sheet and back up again. At this point an interference factor must be applied, and in this case it is zero, thus giving both of these histories zero probability to occur. However if we observe whether or not the photon passed through the sheet on its journey, then in the history where it entered, bounced off the bottom, and then exited through the top again, we will have noticed this and made note of the fact somewhere. In the history where it simply reflects directly off the top, we won't have noticed it, and won't have made any note of the fact anywhere. As these two histories differ in the state of the observer in each case, they don't reconverge to the same overall state, so there is no interference factor, and no interference is observed. Thus any attempt to see whether the photon bounces off the bottom or not destroys interference. This means that the photon's actions upon reaching the bottom boundary won't be biased by considerations of possible interference that could occur in the future, and so this attempt to see into the future will
fail.

This argument is quite general: once we've observed what something is doing, we lose future interference with those histories in which it was doing something else, and so can never use the effect of future interference factors to peer into the future ourselves. Thus although our formulation is acausal, causality violation is always confined to the microscopic world, and cannot be exploited by us.

Some readers may regard the fact that our proposal is acausal as a problem. After all the notion  that the future might affect the present runs contrary to everyday intuition and experience. However such considerations have always proven to be a very poor guide to finding out how the universe works; it is often the case that in realms beyond our everyday experience the laws of physics seem quite bizarre an unnatural. All that really matters is whether the proposal is logically sound and consistent with experiment, and we have shown above that it is. Furthermore acausality is no stranger than various other `unthinkable' notions that exist in conventional quantum physics, such as non-locality and randomness. In fact acausality should be expected of any relativistic quantum theory. We have non-locality in space, and as space and time are mixed up in special relativity under Lorentz transformations, we should expect any theory which has non-locality in space to also have non-locality in time. And just as non-locality in space does not permit faster than light communication, so non-locality in time does not permit one to see into the future.

We should note that this argument assumes that the decoherence process is perfect, however it is not and so there may be a small effect on quantum probabilities at the present time due to future events, but the decoherence process is so close to perfect that it is totally impractical to detect the imperfections. Nevertheless it is conceivable that in the far future technology may advance to the point where causality violations are detectable.

\section{Context and limitations}

In this section we shall briefly compare our theory with some of the more popular alternatives, and discuss some of the possible limitations of the theory.

Over the past fifty years a number of alternatives to the Copenhagen interpretation have been developed. We shall not discuss them in depth here, as the purpose of this paper is to present its theory, rather to act as a review article, however a brief acknowledgement of the main alternatives is appropriate, and the reader who wishes to know more may consult the references.

There are at present a vast number of interpretations/theories of quantum mechanics and it is impossible to mention, or even to know of, them all. However the most popular theories fall into one of the these three classes: Bohmian mechanics type theories \cite{bib:bohm1} \cite{bib:bohm2}, consistent histories \cite{bib:griffiths} \cite{bib:omnes} \cite{bib:gellmann}, or many-worlds \cite{bib:everett}. Our approach does not strictly fall into any of these categories, though it shares a number of concepts and ideas with the many-worlds and consistent histories approaches, while retaining its own distinctive features. All of these approaches have succeeded in depriving observers of any special status in quantum mechanics. However both consistent histories, and the many-worlds theories, are still theories which only work ``for all practical purposes'' (Bell's FAPP) \cite{bib:bell}, in other words they are not based on a well-defined, unambiguous set of laws. Bohmian mechanics is a well-defined theory, but it is non-relativistic, although there have been some recent, partially successful, attempts to construct a relativistic version.

Our theory however, is both well-defined, relativistic, and applicable to the quantum field theories of the standard model. 

There are some points about our theory which may be of some concern. Firstly, the theory is acausal. We have demonstrated in the previous section that this acausality is confined to the microscopic realm, however this is a FAPP argument. In principle, the small variations in quantum probabilities from those predicted by the Copenhagen interpretation, due to the small imperfections of the decoherence process, might be experimentally detectable. Considering the sorts of numbers involved this would appear to be far beyond our current ability to detect, but it is possible that one day causality violations may be experimentally detectable. Although this seems odd, there is nothing illogical about the possibility, as clearly there cannot be time paradoxes in a universe in which an entire history is selected in one go, nor is there any conflict with current observational data as the effect lies far beyond our current ability to detect.

It is also true that the relativistic nature of the theory is one which works only in a FAPP sense. The laws of physics appear relativistic in the special frame of the theory, therefore they do in all frames. However in the special frame there are small fluctuations from the Copenhagen laws of physics, due to the slight imperfections in the decoherence process. There is no reason to suppose that the laws governing these small fluctuations appear relativistic in the special frame, and therefore the laws governing these small fluctuations may be frame dependent. However, as we cannot detect these small fluctuations in the first place, this is also something far beyond our current ability to detect. Thus our proposal results in very small, currently undetectable, violations of causality and special relativity. However there is no logical problem here, nor any conflict with experimental evidence.

Finally we should note that the probabilities for histories may not be well-defined in the limit of the cutoff time $T$ going to infinity, in which case the cutoff time $T$ needs to be specified in order to make the theory well-defined, and it thus has one arbitrary parameter.

\section{Conclusion}

We have in this paper developed a formulation of quantum mechanics based on the concept of selecting particles' histories rather than positions, and then applied it to the whole of spacetime. It has been demonstrated that the usual probability rule from the standard formulation emerges from our theory. Thus it is consistent with experimental evidence, however it does not really predict anything new --- all experimental results must be the same; so it is really just a different way of looking at the same thing, rather than a completely new theory with new results. However it is a formulation which is free of the logical defects of the Copenhagen interpretation. It offers a single view of both the microscopic and macroscopic universe, with no artificial divide between the two. As a related matter, it is also free of the vagueness and ambiguity surrounding the role and nature of observation which afflicts many other formulations; in our formulation observation is not required for the physics to operate smoothly, though we can of course still analyse the effect observations will have if we so choose. Of course it must be said that this is known to be of little practical consequence: people were able to successfully use the old formulation for many decades even without clarity on the nature of observation. Finally it is relativistic, and it appears to be the only proposal which is free of vagueness, free of FAPP, relativistic, and applicable to the quantum field theories of the standard model.

As said above, it has no practical consequences at the present time, nevertheless to bemoan the lack of practical consequences of this formulation is to miss the point. It provides a new perspective on quantum mechanics, and it is always useful to have more than one way of thinking about any physical problem. For example Hamiltonian mechanics was a different way of looking at Newtonian mechanics, but which produced exactly the same experimental predictions. Nevertheless it proved more suitable for analysing certain physical problems, and was thus of great practical use to physicists. Likewise, this formulation of quantum mechanics should prove useful to physicists in various circumstances.

Finally, on the subject of quantum `weirdness', we can see that our formulation does not remove any of the essential weirdness of quantum mechanics. We still have randomness and non-locality, and we have now abandoned causality. What has been removed is the lack of clarity on the dual wave-particle nature of matter, and also the role of the observer has been removed from the basic physics. Our formulation can thus be thought of as one which cleans up quantum theory, but not as one which makes it any less weird.

\end{document}